\pdfoutput=1
\documentclass[aps,floatfix,prd,twocolumn,a4paper,10pt,citeautoscript,
preprintnumbers,longbibliography,superscriptaddress,showpacs]{revtex4-1} 

\usepackage[english]{babel}	

\usepackage{amsmath}		
\usepackage{amssymb}		
\usepackage{bm}			
\usepackage{bbm}			

\usepackage{graphicx}	
\usepackage{graphics}	
\DeclareGraphicsExtensions{.jpg}	

\usepackage[colorlinks=true, pdfstartview=FitV,
			linkcolor=blue, citecolor=blue, urlcolor=blue]{hyperref}

\newcommand{\Equation}[2]{  \begin{equation}\label{#1}#2\end{equation} }
\newcommand{\Align}[2]{\begin{align}\label{#1}#2\end{align}}
\newcommand{\SubAlign}[2]{\begin{subequations}\label{#1}\begin{align}#2\end{align}\end{subequations}}

\newcommand{\Figref}[1]{Fig.~\ref{#1}}
\newcommand{\Eqref}[1]{Eq.\,\eqref{#1}}

\newcommand{\bs}{\boldsymbol}
\newcommand{\groupU}[1]{\mathrm{U}(#1)}   		
\newcommand{\groupZ}[1]{\mathbb{Z}_{#1}} 		
\newcommand{\Exp}[1]{\text{e}^{#1}}
\renewcommand\Re{\mathrm{Re}}
\renewcommand\Im{\mathrm{Im}}

\newcommand{\Grad}{{\bs\nabla}}

\newcommand{\Curl}{{\bs\nabla}\times}

\newcommand{\x}{\bs x}
\newcommand{\F}{\mathcal{F}}
\newcommand{\Hc}[1]{H_{\text{c}#1}} 
\newcommand{\D}{{\bs D}}
\newcommand{\A}{{\bs A}}
\newcommand{\B}{{\bs B}}

\newcommand{\J}{{\bs J}}
\newcommand{\Bext}{{{H}}}

\begin{document}
\title{Lattices of double-quanta vortices and chirality inversion 
in \texorpdfstring{$p_x+ip_y$}{px+ipy} superconductors}

\author{Julien~Garaud}
\email{garaud.phys@gmail.com}
\affiliation{Department of Theoretical Physics and  Center for Quantum Materials, 
KTH-Royal Institute of Technology, Stockholm, SE-10691 Sweden}

\author{Egor~Babaev}				
\affiliation{Department of Theoretical Physics and  Center for Quantum Materials, 
KTH-Royal Institute of Technology, Stockholm, SE-10691 Sweden}

\author{Troels~Arnfred~Bojesen}	
\affiliation{RIKEN Center for Emergent Matter Science, Wako, Saitama, 351-0198, Japan}

\author{Asle~Sudb\o{}}  			
\affiliation{Department of Physics, Norwegian University of Science and Technology, 
NO-7491 Trondheim, Norway}

\date{\today}

\begin{abstract}

We investigate the magnetization processes of a standard Ginzburg-Landau model for 
chiral $p$-wave superconducting states in an applied magnetic field. We find that 
the phase diagram is dominated by triangular lattices of doubly quantized vortices. 
Only in close vicinity to the upper critical field, the lattice starts to dissociate 
into a structure of single-quanta vortices.
The degeneracy between states with opposite chirality is broken in a nonzero field. 
If the magnetization starts with an energetically unfavorable chirality, the process 
of chirality-inversion induced by the external magnetic field results in the formation 
of a sequence of metastable states with characteristic magnetic signatures that can 
be probed by standard experimental techniques. 

\end{abstract}
\maketitle

\section*{Introduction}

The complex structure of the order parameter for chiral $p$-wave superfluid and 
superconducting states has long attracted interest in their physical properties. 
Chiral $p$-wave pairing is realized in the $A$-phase of superfluid $^3$He, 
where the complex structure of the order parameter yields a rich variety of 
topological defects \cite{Mermin.Ho:76,Anderson.Toulouse:77,Walmsley.Golov:12,
Vollhardt.Wolfle,Mermin:79,Volovika}. In the context of superconductivity this 
interest is related to the discovery of Sr$_2$RuO$_4$ \cite{Maeno.Hashimoto.ea:94,
Mackenzie.Maeno:03}, which is argued to have $p$-wave pairing \cite{Rice.Sigrist:95,
Kallin.Berlinsky:09,Kallin:12}, with Cooper pairs having an effective internal 
orbital momentum \cite{Mackenzie.Maeno:03,Nelson.Mao.ea:04}. 

Evidence supporting the existence of a chiral $p$-wave superconducting state in 
Sr$_2$RuO$_4$ has surfaced through a variety of measurements. For instance, the 
superconducting critical temperature ($T_\text{c}$) is completely suppressed by 
adding non-magnetic impurities \cite{Maeno.Hashimoto.ea:94,Mackenzie.Maeno:03}. 
Moreover, NMR Knight shift measurements show no change in the spin susceptibility 
with temperature in the superconducting phase \cite{Murakawa.Ishida.ea:04,
Ishida.Mukuda.ea:98}. Muon spin measurements ($\mu$SR) \cite{Luke.Fudamoto.ea:98} 
and polar Kerr effect \cite{Xia.Maeno.ea:06} suggest that the superconducting state 
breaks time-reversal symmetry. Also, phase-sensitive Josephson spectroscopy 
experiments have shown some evidence of a dynamical domain structure consistent 
with a chiral spin-triplet state \cite{Nelson.Mao.ea:04,Kidwingira.Strand.ea:06}.
Experiments on toroidal mesoscopic samples reporting magnetization with 
half-height steps suggest half-quantum vorticity \cite{Jang.Ferguson.ea:11}, 
while no half-quantum vortices were reported in a singly-connected geometry.

Nevertheless, the nature of the superconducting state of Sr$_2$RuO$_4$ remains 
elusive, since a number of properties predicted for chiral $p$-wave states have 
so far not been observed.
Spontaneous breaking of time-reversal symmetry for chiral $p$-wave state, implies 
the existence of domain-walls (DW) that separate two different time-reversal symmetry 
broken (TRSB) ground states, i.e. different chiral states. As a consequence of 
broken spatial symmetry, these domain walls support spontaneous supercurrents that 
generate magnetic fields \cite{Volovik.Gorkov:85,Sigrist.Rice.ea:89,Matsumoto.Sigrist:99,
Matsumoto.Sigrist:99a,Vadimov.Silaev:13}. Edge currents are also expected to flow 
at the boundaries of samples, quite similarly to the currents at domain walls 
between domains of opposite chirality \cite{Matsumoto.Sigrist:99,Matsumoto.Sigrist:99a,
Vadimov.Silaev:13,Bouhon.Sigrist:14}, and these currents will have a magnetic field 
associated with them. However, in Sr$_2$RuO$_4$, no indication of such a field has 
so far been found in magnetic imaging microscopy experiments \cite{Bjornsson.Maeno.ea:05,
Kirtley.Kallin.ea:07,Hicks.Kirtley.ea:10,Kallin.Berlinsky:09,Curran.Bending.ea:14}. 
Thus, the issue of identifying a possible model of the superconducting state in 
this compound is currently a matter of intense debate  \cite{Raghu.Kapitulnik.ea:10,
Imai.Wakabayashi.ea:12,Raghu.Chung.ea:13,Logoboy.Sonin:09,Scaffidi.Simon:15,
Garaud.Agterberg.ea:12,Bouhon.Sigrist:14,Huang.Scaffidi.ea:16}.

Vortex matter in Sr$_2$RuO$_4$ also shows rich physics that can give insight into 
the nature of superconducting state in this material. The formation of chains of 
vortices has been reported for magnetic fields with an $ab$-plane component 
\cite{Dolocan.Lejay.ea:06}, consistent with the mechanism of vortex chain formation 
in layered systems. 
Small-angle neutron scattering \cite{Riseman.Kealey.ea:98}, and muon-spin rotation 
measurements \cite{Aegerter.Lloyd.ea:98,Ray.Gibbs.ea:14}, have revealed vortex lattices 
with square symmetry at high fields. A transition to triangular a vortex lattice at lower 
fields has been reported in \cite{Curran.Khotkevych.ea:11,Ray.Gibbs.ea:14}. Such transitions 
of the vortex lattice structure have been regarded as being consistent with predictions based 
on lowest-Landau-level calculations for chiral $p$-wave superconductivity in Sr$_2$RuO$_4$ 
\cite{Agterberg:98,Agterberg:98a,Heeb.Agterberg:99}. However, they are inconsistent with 
numerical studies of the energy of isolated topological defects \cite{Garaud.Babaev:15a} 
that have predicted the formation of double-quanta vortices in the Ginzburg-Landau model for 
a chiral $p$-wave superconductor. Early experiments also demonstrated ``zero creep'' that 
is not accompanied by a dramatic rise in critical current \cite{Dumont.Mota:02}. This indicates 
that vortices form relatively mobile clusters. The initial interpretation \cite{Dumont.Mota:02} 
of this experiment was taken as evidence for a chiral $p$-wave state that allows the formation 
of groups of type-2 vortices trapped by a closed chiral domain wall. Within this scenario 
the domain wall would prevent vortex creep outside the sample. At the same time, in contrast 
to the vortex pinning scenario, these groups of vortices could be moved by an external current. 
This would explain the absence of a dramatic rise in the critical current. However, such a 
configuration would have characteristic magnetic signatures (see Refs. \cite{Garaud.Babaev:12,
Garaud.Babaev:15a} and discussion below). These signatures have not been seen so far in 
scanning surface probes. Instead, experiments using magnetic surface probes have reported 
observations of clusters of integer-flux vortices \cite{Dolocan.Veauvy.ea:05,
Bjornsson.Maeno.ea:05,Hicks.Kirtley.ea:10}. Evidence of vortex clustering has also 
been found in bulk measurements in field-cooled muon-spin rotation experiments 
\cite{Ray.Gibbs.ea:14}. The key observation there was that vortex clusters contract as 
temperature is lowered well below $T_c$, which is inconsistent with vortex pinning. 
Ref.~\cite{Ray.Gibbs.ea:14} has attributed vortex coalescence to the competition between 
multiple coherence lengths that may originate from multi-band effects or other multicomponent 
order parameters of various origins (such a ``type-1.5" scenario was hypothesized in an 
earlier paper \cite{Hicks.Kirtley.ea:10} in analogy with \cite{Moshchalkov.Menghini.ea:09}).

In zero field, both chiral (ground) states are degenerate in energy, while this 
degeneracy is lifted by a magnetic field. For a given orientation of the magnetic 
field, only one of the chiral states is stable while the time-reversed chiral 
state is energetically penalized. Hence, the dominant component can form a vortex. 
Since the dominant component is suppressed in the vicinity of the vortex core, 
the time-reversed (subdominant) chiral component may be induced in the vortex 
core \cite{Heeb.Agterberg:99,Ichioka.Machida:02}.
The winding of the induced component is not independent of that of the dominant 
component. It has a $4\pi$ winding of the relative phases that follows from the Cooper 
pairs having nonzero internal orbital momentum \cite{Sauls:94}. Since the magnetic 
field lifts the degeneracy between chiralities, vortices with opposite phase 
winding have different physical properties \cite{Ichioka.Machida:02,Sauls.Eschrig:09,
Garaud.Babaev:15a}.

Apart from single-quanta vortices, there also exist stable vortices carrying multiple 
quanta of magnetic flux. These are essentially different from single-quanta vortices, 
because as they are coreless they carry an additional topological charge, and they are 
sometimes called skyrmions. As discussed in more detail below, the component induced 
by a doubly quantized vortex in the dominant component has zero winding 
\cite{Ichioka.Machida:02,Sauls.Eschrig:09}. The possible existence of lattices of 
double-quanta vortices has been proposed earlier in the context of the heavy fermion 
compound UPt$_3$ \cite{Barash.MelNikov:90,Melnikov:92}, which is believed to be 
described by a similar type of model \cite{Joynt:91,Joynt.Taillefer:02}. 
Based on self consistent calculations using Eilenberger theory for the chiral $p$-wave 
state, it was recently argued that while lattices of single-quanta vortices form for
fields close to $\Hc{1}$, lattices of double-quanta vortices are favored in higher 
fields \cite{Ichioka.Machida.ea:12}. On the other hand, within the Ginzburg-Landau theory 
for chiral $p$-wave superconductors, double-quanta (coreless) vortices have been 
shown to be energetically favored as compared to two (isolated) single-quanta vortices 
\cite{Garaud.Babaev:15a} and they were also found to appear in a mesoscopic sample 
\cite{FernandezBecerra.Sardella.ea:16}. The energetic preference for double-quanta vortices 
does not exclude the formation of lattices of single-quantum vortices, or more complicated 
structures in a magnetization process. Interactions can favor different vortex lattices, 
or different Bean-Livingston barriers may result in the formation of metastable lattices 
for vortices that are not the most energetically favorable. This raises the question 
of the nature of magnetization processes and what kind of lattices form when an external 
magnetic field is applied.

In this paper, we investigate magnetization processes, using numerical simulations of 
the minimal Ginzburg-Landau theory describing the chiral $p$-wave state in an external 
field directed along the ${\bf c}$ axis. In Section \ref{Sec:Theory}, we introduce the 
Ginzburg-Landau theory used to describe the chiral $p$-wave state in an external field, 
and we discuss various basic properties, such as ground states and edge currents. Next, 
Sec.~\ref{Sec:double-quanta-lattices} is devoted to the magnetization process that has 
minimal energy, i.e. when the external field produces topological excitations with lowest 
energy. In that case, we find that lattices of double-quanta vortices are generically 
produced. Finally, Sec.~\ref{Sec:Chirality-switching} investigates the magnetization 
processes with reversed magnetic field. These states have higher energies and eventually 
lead to chirality inversion via a subtle interplay between vortices and domain walls.

\section{Ginzburg-Landau Model}
\label{Sec:Theory}

In the coordinate system in which the crystal anisotropy axis is ${\bf c}\parallel{\bf z}$, 
the $p_x+ip_y$ state corresponds to the two-dimensional representation 
$\Gamma_5^-=(k_x{\bf z},k_y{\bf z})$ and the order parameter is described by a 
two-dimensional complex vector ${\bs\eta}=(\eta_x,\eta_y)/\sqrt{2}$ \cite{Mackenzie.Maeno:03,
Joynt.Taillefer:02,Sigrist.Ueda:91}. Introducing the chiral order parameter basis 
$\eta_\pm=\eta_x\pm i\eta_y$, the dimensionless Ginzburg-Landau free energy reads as 
(see e.g. \cite{Agterberg:98,Agterberg:98a,Heeb.Agterberg:99}): 
\SubAlign{Eq:FreeEnergy}{
&\mathcal{F}= 
	|\Curl\A|^2 + |\D\eta_+|^2 + |\D\eta_-|^2 
	\label{Eq:GradientEnergy1} \\
   &+(\nu+1)\Re\left[ (D_x\eta_+)^*D_x\eta_- - (D_y\eta_+)^*D_y\eta_- \right]	
    \label{Eq:GradientEnergy2} \\
   &+(\nu-1)\Im\left[ (D_x\eta_+)^*D_y\eta_- + (D_y\eta_+)^*D_x\eta_- \right]	
    \label{Eq:GradientEnergy3} \\
   &+2|\eta_+\eta_-|^2+\nu\Re\left(\eta_+^{*2}\eta_-^2\right)
   	+\sum_{a=\pm}-|\eta_a|^2+\frac{1}{2}|\eta_a|^4   
	\label{Eq:PotentialEnergy}	.
}
Here $\eta_\pm=|\eta_\pm|\Exp{i\varphi_\pm}$ and we have used dimensionless units were 
the free energy is normalized to the condensation energy, and the lengths are given in 
units of $\xi=\left(\alpha_0(T-T_\text{c})\right)^{-1/2}$. The magnetic field $\B=\Curl\A$ 
is given in units of $\sqrt{2}B_\text{c}=\Phi_0/(2\pi\lambda\xi)$. 
The dimensionless gauge coupling $g$ that appears in the covariant derivative 
$\D=\Grad+ig\A$ is used to  parametrize the ratio of two length scales in this 
Ginzburg-Landau model, $g^{-1}:=\kappa=\lambda/\xi$. The anisotropy parameter $\nu$, 
which satisfies $|\nu|<1$, determines the anisotropy in the $xy$-plane. It measures 
the tetragonal distortions of the Fermi surface, which has cylindrical geometry for 
$\nu=0$, and it is defined as $\nu=(\langle v_x^4\rangle-3\langle v_x^2v_y^2\rangle)/
(\langle v_x^4\rangle+\langle v_x^2v_y^2\rangle)$ (where $\langle\cdot\rangle$ denotes 
the average over the Fermi surface). In the model defined by \Eqref{Eq:FreeEnergy}, 
the dependence on the third coordinate is not considered (i.e. assuming a two-dimensional 
system or translational invariance along the $z$-axis). 
Varying \Eqref{Eq:FreeEnergy} with respect to $\eta_\pm$ yields the Ginzburg-Landau 
equations given by
\Align{Eq:GLequations}{
&\Pi_{x^2+y^2}\eta_\pm
+\left(\frac{\nu+1}{2}\Pi_{x^2-y^2}\pm\frac{\nu-1}{2i}\Pi_{xy} \right)\eta_\mp
=\frac{\partial\F_p}{\partial\eta_\pm^*}	\nonumber\\
&\text{with}~\Pi_{x^2\pm y^2}=D_xD_x\pm D_yD_y\,,~
\Pi_{xy}=\left\{D_x,D_y\right\}\,,
}
where $\F_p$ is the potential term \Eqref{Eq:PotentialEnergy} in the free energy and 
$\{\cdot,\cdot\}$ stands for the anti-commutator. Variation with respect to the vector 
potential gives Amp\`ere's equation $\Curl\B+\J=0$, where the total current is the sum 
of partial currents $\J^\pm$ whose components are 
\SubAlign{Eq:Current}{
J^\pm_x=\frac{g}{2}&\Im\Bigg(
\eta^*_\pm\Big( D_x\eta_\pm
+\phantom{i}\left[D_\pm +\nu D_\mp \right]\frac{\eta_\mp}{2}
\Big)\Bigg)\,,\\
J^\pm_y=\frac{g}{2}&\Im\Bigg(
\eta^*_\pm\Big( D_y\eta_\pm
\pm i\left[D_\pm -\nu D_\mp \right]\frac{\eta_\mp}{2}
\Big)\Bigg)\,,
}
where $D_\pm=D_x\pm iD_y$.

The theory described by \Eqref{Eq:FreeEnergy} has several symmetries. Firstly, 
\Eqref{Eq:FreeEnergy} exhibits the usual $\groupU{1}$ gauge invariance under the 
transformation ${\bs\eta}\to \Exp{i\zeta(\x)}{\bs\eta}$ and $\A\to\A-\Grad\zeta(\x)/g$. 
The theory is also invariant under a discrete ($\groupZ{2}$) operation $\mathcal{T}$, 
which is referred to as time-reversal symmetry, 
$\{\eta_\pm,\B\}\xrightarrow{\mathcal{T}}\{\eta^*_\mp,-\B\}$. As discussed below, 
the chiral ground state spontaneously breaks this symmetry.

The nontrivial behavior of the superconducting degrees of freedom at the boundary of 
a sample is responsible for the generation of spontaneous edge currents. Within the 
Ginzburg-Landau theory, this behavior is accounted for by adding relevant surface 
terms of the form \cite{Sigrist.Ueda:91}: 
\Align{Eq:FreeEnergySurface}{
\F_\text{surf}&=\left[\chi_1(n_x^2+n_y^2)+\chi_z n^2_z\right]
	\left[|\eta_+|^2+|\eta_-|^2 \right]  	 \\
	&+2\chi_2(n_x^2-n_y^2)\Re\left(\eta_+^*\eta_-\right)
	-4\chi_3n_xn_y\Im\left(\eta_+^*\eta_-\right)\,, \nonumber
}
where $n_i\equiv{\bs n}\cdot {\bs{\hat \imath}}$ are the components of the normal vector 
to the boundary. In our two-dimensional problem $\chi_z=0$ and for simplicity, we can 
choose $\chi_1=\chi_2=\chi_3\equiv\chi$, imposing specular reflection on the boundary. 
The magnetization processes are thus described by the total (Gibbs) free energy over 
a domain $\Omega$,
\Equation{Eq:GibbsFreeEnergy}{
{G}=\int_\Omega\F-2\B\cdot\Bext{\bs{\hat z}}+\int_{\partial\Omega}\F_\text{surf}\,, 
}
together with the conditions that $\Curl\A=\Bext{\bs{\hat z}}$ at the boundary 
$\partial\Omega$ of the domain. Here $\Bext$ denotes the strength of an external 
field.

\subsection{Details of the numerics}
\label{Sec:Numerics}

To numerically minimize \Eqref{Eq:GibbsFreeEnergy}, the physical degrees 
of freedom $\eta_\pm$ and $\A$ are discretized using a finite-element framework 
\cite{Hecht:12,Hutton,Reddy}. First we construct a mesh being a regular partition 
of the spatial domain using a Deleaunay-Voronoi triangulation algorithm. In other 
words, the spatial domain is subdivided into as set of triangles (having similar area). 
Then, $\eta_\pm$ and $\A$ are expressed in terms of second order Lagrange polynomials 
(polynomials of $x,y$ up to second order) on each triangle. This means that on a given 
triangle, each of the six physical degrees of freedom of the problem ($\eta_\pm$, 
$\eta_\pm^*$, and $\A$) is parametrized by the six coefficients of the second-order 
interpolating polynomials (there are six independent coefficients for a second-order 
polynomial in two dimensions). The second order Lagrange interpolation defines the 
six coefficients at vertices and mid-edges, for a total of $6\times6=36$ numerical 
degrees of freedom per triangle. The overall accuracy of the construction is determined 
by the number of triangles that constitute the mesh, as well as the order of the 
interpolation method.

Now, within this finite element framework, we use a nonlinear conjugate-gradient algorithm 
(see, e.g., Ref.~\onlinecite{Nocedal.Wright}), which is iterated until relative variations 
of the norm of the functional gradient with respect to all degrees of freedom is less 
than $10^{-8}$.

In this work we investigate magnetization processes and vortex structure formation 
due to an applied magnetic field on domains of finite size. We focus on characteristic 
states that appear during magnetization processes, and which should be experimentally 
observable. We therefore do not specifically focus on the question of which vortex 
lattice is a ground state in a given field in the thermodynamic limit. Precise answers 
to minimal energy structure in a thermodynamic limit would require a different approach. 
There are intrinsic limitations to characterize a lattice structure, when working on 
finite domains. 
First of all, realizing perfectly ordered lattices typically requires a certain number 
of vortices given a certain area. Unfortunately, during magnetization processes, the 
number of vortices varies, and hence the appropriate number of vortices may not be 
realized.
Moreover, unlike in periodic domains, the overall lattice structure is determined 
by more than just intervortex forces. The existence of Meissner currents flowing along 
boundaries can also alter the lattice structure. 
Although such effects should tend to be less important in very large domains, this 
explains why, in rather high fields the structure we find can be distorted or less 
ordered.

\subsection{Ground-state}

The ground-state that minimizes the potential energy, \Eqref{Eq:PotentialEnergy}, is 
degenerate and the solutions are $(\eta_+,\eta_-)=(1,0)$ and $(0,1)$. Symmetrywise it 
spontaneously breaks the $\groupU{1}\times\groupZ{2}$ symmetry, where $\groupZ{2}$ 
refers to time-reversal operations. The spontaneous breakdown of the discrete $\groupZ{2}$ 
symmetry dictates that the theory allows domain wall solutions that interpolate 
between regions in different ground states. Such domain walls carry 
a magnetic field perpendicular to the $xy$-plane \cite{Matsumoto.Sigrist:99,
Matsumoto.Sigrist:99a}. Aspects of the domain wall physics and their role in 
chirality switching are discussed later, in Sec.~\ref{Sec:Chirality-switching}.

The discrete ($\groupZ{2}$) degeneracy of the ground state is lifted for a nonzero 
applied field $H\hat{z}$. Consider for example a constant magnetic field induced 
by the external field $\B=B_z\hat{z}=H\hat{z}$, if $B_z>0$ the ground state is 
$(\eta_+,\eta_-)=(1,0)$, while when $B_z<0$ the lowest energy state is 
$(\eta_+,\eta_-)=(0,1)$. As the $\eta_+$ and $\eta_-$ components behave differently 
in external field, a complete study for a given ground-state necessitates considering 
both situations $B_z>0$ and $B_z<0$. Note that due to the time-reversal symmetry of 
the theory $\{\eta_\pm,\B\}\xrightarrow{\mathcal{T}}\{\eta^*_\mp,-\B\}$, this is 
equivalent to investigating only a fixed direction of the magnetic field (say $B_z>0$), 
however including both chiral states. 
In the following, we choose to fix the dominant component of the order parameter to 
be $\eta_-$ (i.e. the ground state is $(\eta_+,\eta_-)=(0,1)$) and thus investigate 
both positive and negative applied magnetic field. 

\subsection{Edge currents}

Spontaneous currents are expected to appear at the boundaries of chiral $p$-wave 
superconducting samples. However, scanning Hall \cite{Bjornsson.Maeno.ea:05} and 
scanning SQUID microscopy \cite{Kirtley.Kallin.ea:07,Hicks.Kirtley.ea:10} experiments 
in Sr$_2$RuO$_4$ have not detected such predicted edge currents, which
in general should affect magnetization processes of chiral $p$-wave superconductors. 
If such edge currents are strong enough, the physics of vortex entry into the system
can be substantially modified compared to that in ordinary superconductors. 
Indeed, as discussed in detail below, the edge current can affect the 
Bean-Livingston barrier and hence, the processes of vortex entry. For example 
it can either facilitate or suppress vortex entry near $\Hc{1}$, a fact which 
will affect the chirality inversion process.

\begin{figure}[!htb]
\hbox to \linewidth{ \hss
\includegraphics[width=\linewidth]{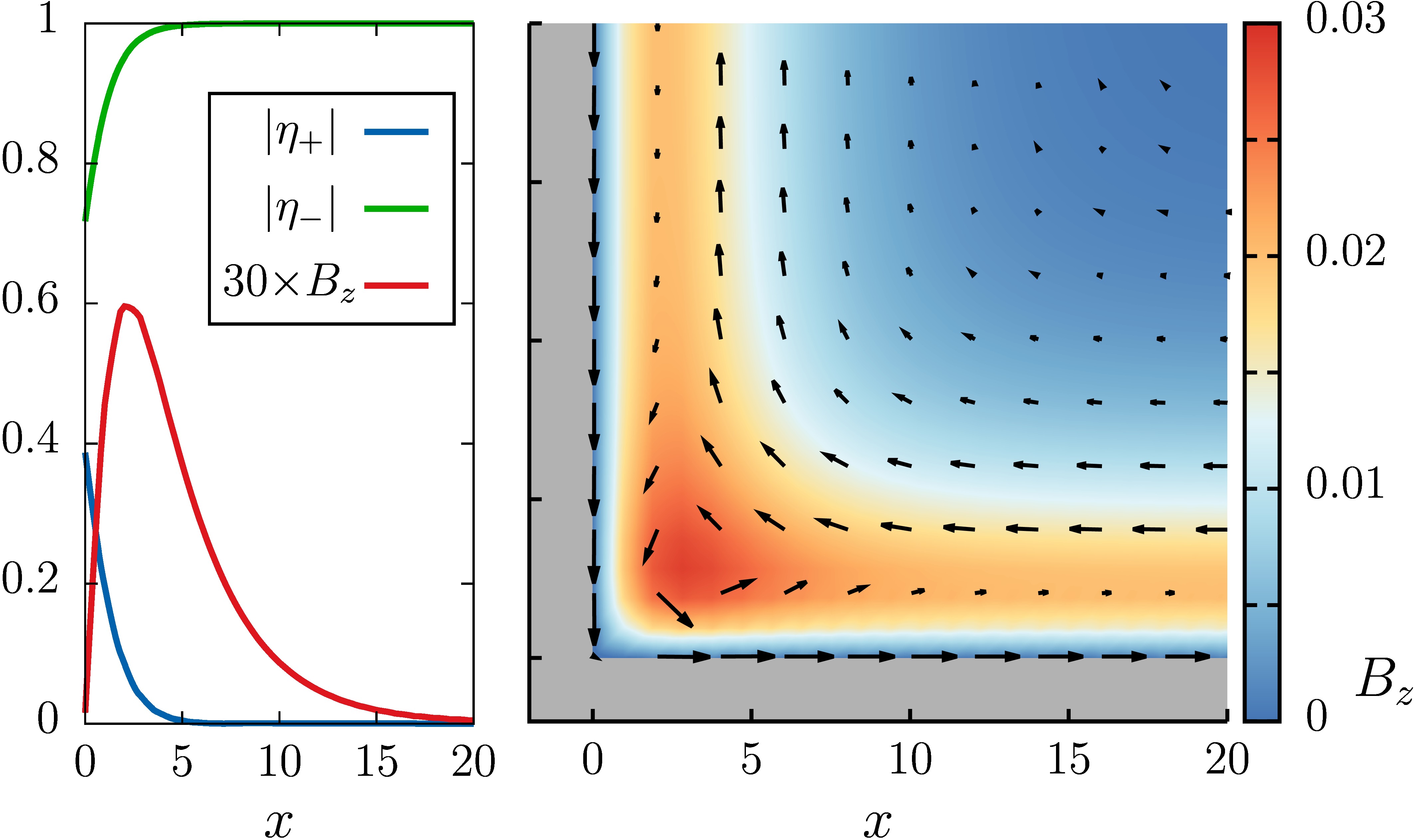}
\hss}
\caption{
(Color online) -- 
Properties of the edge currents due to the surface term \Eqref{Eq:FreeEnergySurface} 
for $\chi_1=\chi_2=\chi_3=1$. Left panel shows the behavior of the components 
$|\eta_\pm|$ and the magnetic field as a function of the distance from a straight 
edge boundary along the $y$-axis.
The right panel shows the circulating edge current and the induced magnetic field 
near a corner.
}
\label{Fig:EdgeCurrents}
\end{figure}

The spontaneous magnetic field due to edge currents is found by minimizing 
\Eqref{Eq:GibbsFreeEnergy} in zero external field ($\Bext=0$). Figure~\ref{Fig:EdgeCurrents} 
shows that the spontaneous currents at the edges (here circulating counter-clockwise) 
induce a magnetic field that is screened in the bulk by superconducting currents 
(here circulating clockwise). The calculation clearly shows that the magnetic field 
in the corner is enhanced as compared to a straight edge.

Note that the orientation of the edge currents is specified by the chirality of the 
superconducting state. For example, in \Figref{Fig:EdgeCurrents} the dominant component 
is $\eta_-$ and the currents circulate counter-clockwise. In the case in which the 
dominant component is $\eta_+$, the currents circulate clockwise.
In principle, the surface term, \Eqref{Eq:FreeEnergySurface}, responsible for the edge 
currents, should not affect the bulk properties, such as, for example, vortex lattices. 
However, as discussed below, since the surface term modifies the boundary behavior, it 
can strongly influence vortex entry during a magnetization process and thus lead to 
qualitatively new features. In general, the boundary terms are important 
at low fields and have less influence at high fields.

\section{Lattices of double-quanta vortices}
\label{Sec:double-quanta-lattices}

As stated above, the discrete degeneracy of the chiral ground state is lifted by 
an external field. This implies that given a ground state (which we take to be 
$(\eta_+,\eta_-)=(0,1)$), the magnetization processes will be different whether 
the applied field is parallel or anti-parallel to the ${\bf c}$-axis. Similarly, 
vortices with counter-clockwise winding have different energy than vortices 
with clockwise winding.
After briefly reviewing the elementary properties of vortex matter in the theory 
of a chiral $p$-wave superconducting state, \Eqref{Eq:FreeEnergy}, we investigate 
the magnetization processes when $\Bext<0$, i.e. the case when an applied field 
excites vortices which have least energy. This magnetization process is that of 
least energy, and it exhibits the formation of a triangular lattice of double-quanta 
vortices, which dissociates into a lattice of single-quanta vortices in the 
vicinity of the upper critical field $\Hc{2}$.

\subsection{Isolated vortices and skyrmions}

The asymptotic vorticity of the dominant component $\eta_-$ determines the 
sign of $B_z$, as well as the vorticity of the subdominant component $\eta_+$ 
\cite{Heeb.Agterberg:99}, according to:
\Equation{Eq:vorticity}{
\eta_-\propto\Exp{in_-\theta}\,,~
\eta_+\propto\Exp{in_+\theta} 
~~~\text{and}~~~n_+=n_-+2~\in \mathbb{Z}\,.
}
The relative phase $\varphi_--\varphi_+$ between the components $\eta_+$ and 
$\eta_-$, that corresponds to a difference $\Delta l=2$ of the order parameters' 
angular momentum, originates with the structure of mixed gradients, 
Eqs.~\eqref{Eq:GradientEnergy2} and \eqref{Eq:GradientEnergy3}. Note that since 
the subdominant component, $\eta_+$, vanishes asymptotically (i.e., it recovers its 
ground state value $\eta_+=0$ in the bulk phase), the winding $n_+$ can be located 
only in the close vicinity of a vortex core. Hence, the number of flux quanta is 
determined only by the winding number $n_-$ of the dominant component. 
\Eqref{Eq:vorticity} implies that the two possible single-quanta vortices are 
$(n_-,n_+)=(+1,+3)$ and $(n_-,n_+)=(-1,+1)$. Having different winding numbers of 
the subdominant component, these will have different core structures and it is 
thus natural to expect that they will have different energies as well.

In agreement with the naive expectation, since it has a simpler core structure, 
the $(n_-,n_+)=(-1,+1)$ vortex can have a lower energy than the $(n_-,n_+)=(+1,+3)$ 
vortex \cite{Sauls.Eschrig:09,Garaud.Babaev:15a}. As a result, the vortex with the 
lowest energy carries magnetic field anti-parallel to the ${\bf c}$-axis (for the 
case where the dominant component is $\eta_+$, the lowest energy vortex carries a 
magnetic field parallel to the ${\bf c}$-axis). 
The preference for the $(n_-,n_+)=(-1,+1)$ vortex, featuring the simpler core structure, 
occurs in the whole $(\nu,g)$ parameter space (at least within the Ginzburg-Landau 
model, \Eqref{Eq:FreeEnergy}) \cite{Garaud.Babaev:15a}. It also follows that 
$(n_-,n_+)=(-1,+1)$ and $(n_-,n_+)=(+1,+3)$ have different lower critical fields, 
$\Hc{1}^{(-1)}<\Hc{1}^{(+1)}$. In other words, given a dominant component in the 
ground state, the first vortex entry occurs at different values of the applied 
field, according to if it is parallel or anti-parallel to the ${\bf c}$-axis 
\cite{Matsunaga.Ichioka.ea:04,Ichioka.Matsunaga.ea:05}.

\begin{figure}[!htb]
\hbox to \linewidth{ \hss
\includegraphics[width=0.95\linewidth]{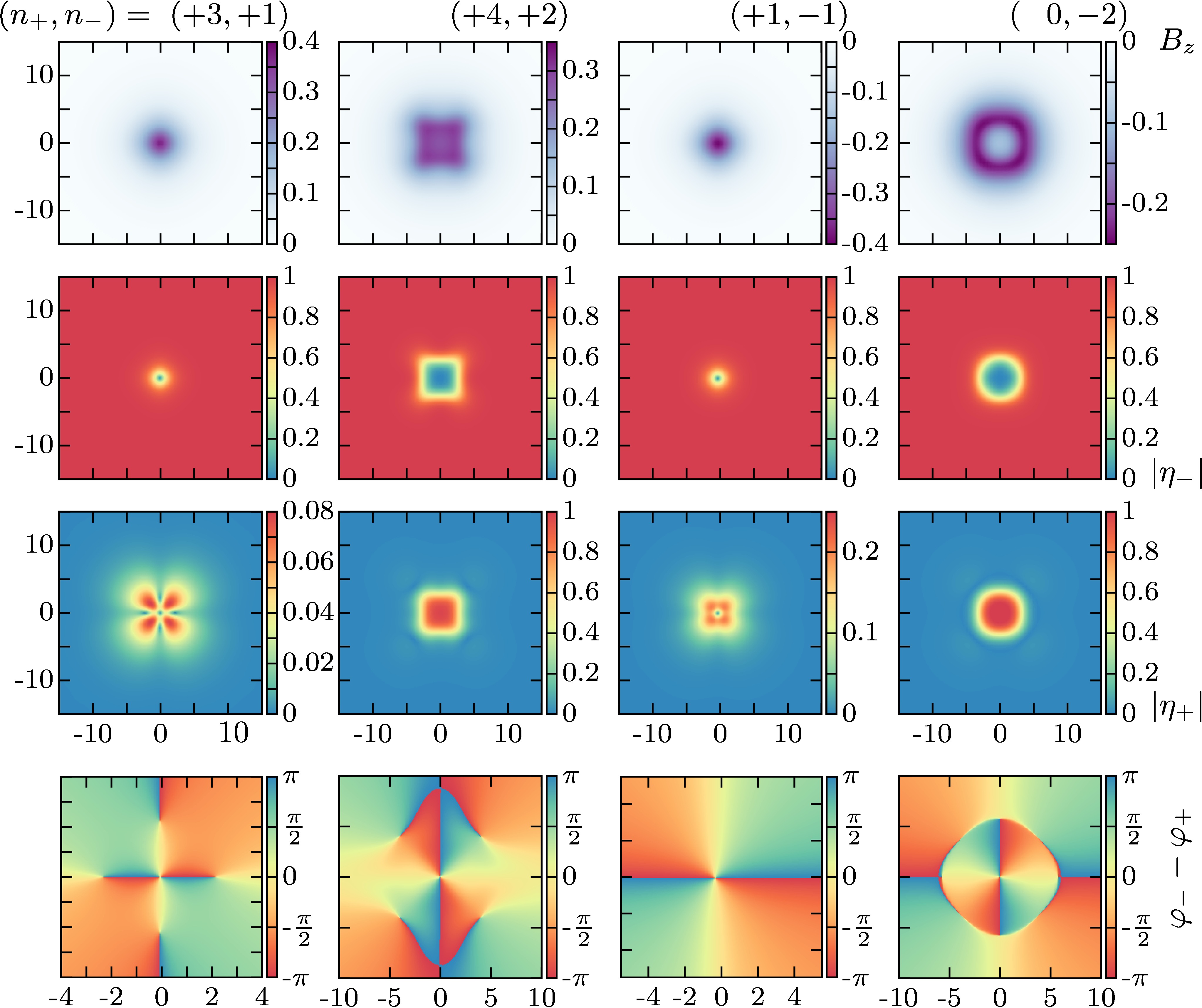}
\hss}
\caption{
(Color online) -- 
Vortex states for the model defined in \Eqref{Eq:FreeEnergy} and \Eqref{Eq:FreeEnergySurface} 
with parameters $g=0.3$ and $\nu=0.2$. The first line shows the magnetic field $\B$, while 
the second and third lines display $\eta_-$ and $\eta_+$, respectively. The fourth line shows 
the relative phase $\varphi_--\varphi_+$ between $\eta_+$ and $\eta_-$. Winding the relative 
phase indicates the position of the cores of $\eta_+$ and $\eta_-$. The first two columns 
show respectively single and double-quanta vortices with $B_z>0$, while the third and fourth 
columns display single and double-quanta vortices with $B_z<0$. 
}
\label{Fig:Vortices}
\end{figure}

The winding number $n_-$ of the dominant component $\eta_-$ specifies the topological 
sector. In infinite domains, different topological sectors are separated by an infinite 
energy barrier, which becomes finite (but still very high) in finite spatial domains. 
This implies a `topological protection' because no continuous finite-energy transformation 
can change the topological sector. As a result, a minimization algorithm that continuously 
deforms the field configurations to reduce the energy cannot change the number of flux 
quanta 
\footnote{
Note that this is rigorously true in infinite domains, while in finite domains there is 
the possibility to change the topological sector by entering/exiting topological defects 
(vortices) through the boundary of the domain. 
}. 
More precisely, starting with a configuration having a given winding $n_-$, the specifics 
of the minimization algorithm can affect core structures, but the asymptotic behavior of 
the vortices after convergence of the algorithm, will, regardless of the algorithmic details,
naturally behave as expected from \Eqref{Eq:vorticity}. 
The heuristic argument of the simplicity of the core structure of the vortices also 
implies the rather unusual situation that double-quanta vortices could be favored 
compared to two isolated single-quanta vortices \cite{Ichioka.Machida:02,Sauls.Eschrig:09,
Ichioka.Machida.ea:12,Garaud.Babaev:15a,FernandezBecerra.Sardella.ea:16}. Indeed, the 
double-quanta $(n_-,n_+)=(-2,0)$ vortex, as it has a simple core structure, is always 
energetically favored  compared to two isolated single-quanta $(n_-,n_+)=(-1,+1)$ 
vortices \cite{Garaud.Babaev:15a}. Thus, one may expect the double-quanta vortices to 
form in an external field, at least close to $\Hc{1}$. However, as they carry more flux, 
their entry through the boundary could be unlikely because they experience a different 
Bean-Livingston barrier.

\begin{figure*}[!htb]
\hbox to \linewidth{ \hss
\includegraphics[width=.9\linewidth]{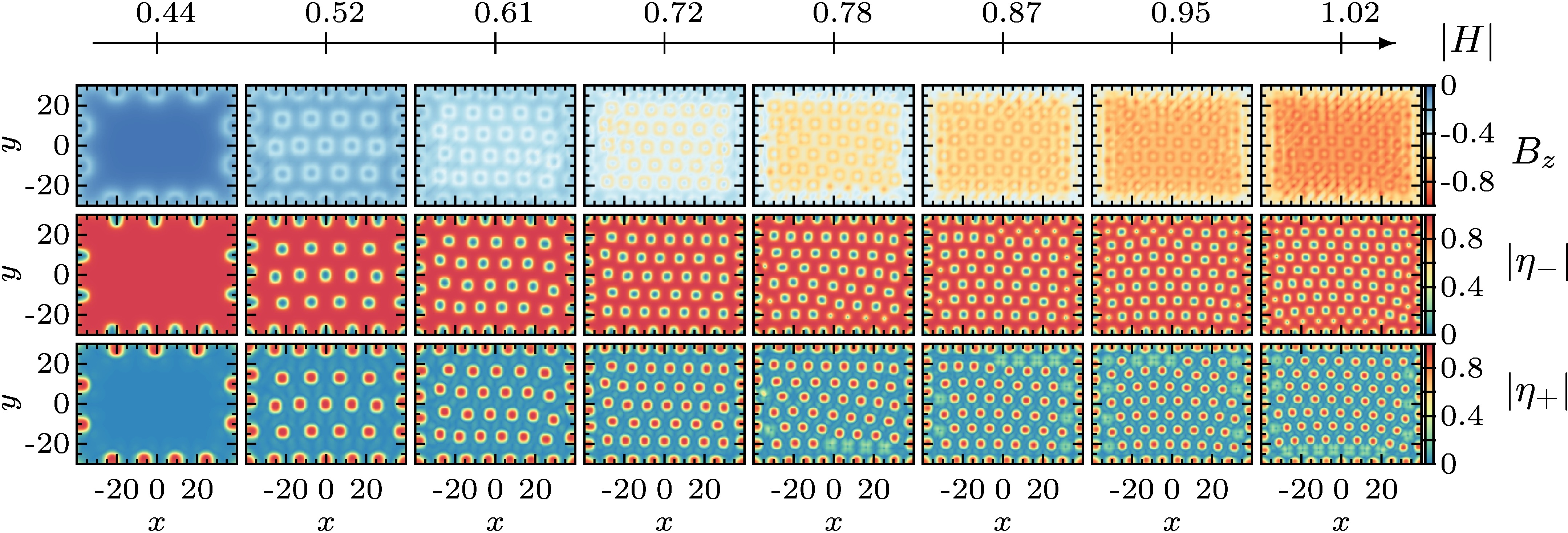}
\hss}
\caption{
(Color online) -- 
Simulation in an external field $\Bext<0$ for the parameters $g=0.3$ and $\nu=0.3$ 
and $\chi=0.1$. The different lines display $B_z$, $|\eta_-|$, and $|\eta_+|$, 
respectively. Here, the orientation of the external field is such that it will 
produce less energetic topological defects (skyrmions). 
The panel corresponding to the lower field show half-quanta vortices stabilized near 
boundary. Increasing the field past $\Hc{1}$  produces entry of double-quanta vortices 
that arrange themselves into a hexagonal lattice.
Note that in higher fields, both single- and double-quanta vortices enter the system. 
The single-quanta vortices will eventually merge into double-quanta skyrmions.
}
\label{Fig:RampUp1}
\end{figure*}

\Figref{Fig:Vortices} illustrates the richness of the core structure of vortices. 
It is evident that vortices with opposite winding of the dominant component 
$n_-=\pm1$ have different structures. In particular, the position and number 
of cores of the different components can be extracted from the last row, which 
shows the relative phase $\varphi_--\varphi_+$. Far away from the cores, 
the components reach their asymptotic values $\varphi_\pm = n_\pm \theta$ 
given by \Eqref{Eq:vorticity}. Thus, the relative phase $\varphi_--\varphi_+$ 
shows the expected $4\pi$ winding at large distances.
\Figref{Fig:Vortices} also displays the two possible configurations carrying two 
flux quanta. Clearly, these also have different core structures. The $(n_-,n_+)=(-2,0)$ 
vortices are always less energetic than two isolated single-quanta vortices with 
$(n_-,n_+)=(-1,+1)$ (see Ref.~\cite{Garaud.Babaev:15a} for a detailed analysis).
Note that there also exist $(n_-,n_+)=(+2,+4)$ vortices. Their energy, compared 
to that of isolated $(n_-,n_+)=(+1,+3)$ vortices, can either be larger or smaller 
depending on the parameters $(\nu,g)$. In the regimes investigated here, the 
double-quanta $(n_-,n_+)=(+2,+4)$ vortices have higher energy than isolated ones. 
Thus, they are only meta-stable. 
Alternatively, the vortices discussed above can be understood as bound states of 
half-quantum vortices in term of the components $(\eta_x,\eta_y)$ of the order 
parameter (see the corresponding discussion in 
Appendix~\ref{App:VorticesPXPY}).
These (coreless) vortices carrying multiple flux-quanta can be characterized by 
additional topological invariants, motivating the alternate terminology of skyrmions 
\cite{Garaud.Babaev:15a}.

\subsection{Magnetization process -- Lattices of double-quanta vortices}

The physics of isolated vortices strongly suggests that double-quanta vortices 
should form in an external field. Here, we investigate the magnetization processes, 
starting from the Meissner state and ramping-up the applied field anti-parallel 
to the ${\bf c}$-axis ($\Bext<0$). The solution in zero field is chosen to be the 
$(\eta_+,\eta_-)=(0,1)$ ground state. The external field is then sequentially 
increased (in steps of $4\times10^{-3}$), and the energy is minimized at each step. 
Figure~\ref{Fig:RampUp1} shows the outcome of such a magnetization process.
This procedure corresponds to an applied field that, in the sense that it produces 
the less energetic defects, is optimally directed. As expected from the properties 
of the isolated vortices, the initial vortex entry comes in the form of double-quanta 
vortices. In our simulation, \Figref{Fig:RampUp1}, the initial entry occurs 
at $|\Bext|\simeq0.46$.
As the applied field increases, more double-quanta vortices enter and they arrange 
themselves in a regular lattice of double-quanta skyrmions. This lattice state is 
robust and persists for all applied fields. The preference for lattices of double-quanta 
vortices, in the case of an anti-parallel external field, is a robust feature. We  
observed this behavior for all the parameters of the model we considered. 

Since the strength of the edge currents depends on $\chi$, the Bean-Livingston barrier 
for vortex entry is affected as well. We find that the entry of skyrmions occurs for 
a wide range of values of the parameter $\chi$ that parametrizes the edge properties. 
The value of the field for initial entry depends on the interplay with the edge currents. 
Nonetheless, bulk properties are essentially unaffected such that lattices of 
double-quanta vortices are always realized.

As stated earlier in more details in Sec.~\ref{Sec:Numerics}, characterizing the lattice 
structure, within our framework of working on a finite size domain can be difficult. 
Due to uncontrollable vortex entry during the magnetization process and the interaction 
between vortices and Meissner currents flowing along the edge of the domain, perfect 
lattice structures are in practice never realized. 
Nonetheless, at least in rather low fields, it is quite clear that hexagonal lattices 
of two-quanta vortices are realized. In higher fields, the coexistence of a few 
single-quanta vortices distorts the overall structure, but the tendency to form hexagonal 
lattice is nevertheless quite robust.

\subsection{Lattice dissociation near \texorpdfstring{$\Hc{2}$}{Hc2}}

The results in \Figref{Fig:RampUp1} show the magnetization process from low to 
rather high fields, when $\Bext<0$ is optimally directed. It is important to 
further understand the behavior in high fields near the second critical field 
$\Hc{2}$. From the energetics of isolated vortices and from the magnetization 
processes, one would conclude that the double-quanta vortices are always 
favored for the model we consider. This would contradict earlier calculations 
using lowest Landau-levels-based approach predicting that a square lattice of 
single-quanta vortices is the solution near $\Hc{2}$ \cite{Agterberg:98,Agterberg:98a}.

\begin{figure*}[!tb]
\hbox to \linewidth{ \hss
\includegraphics[width=.65\linewidth]{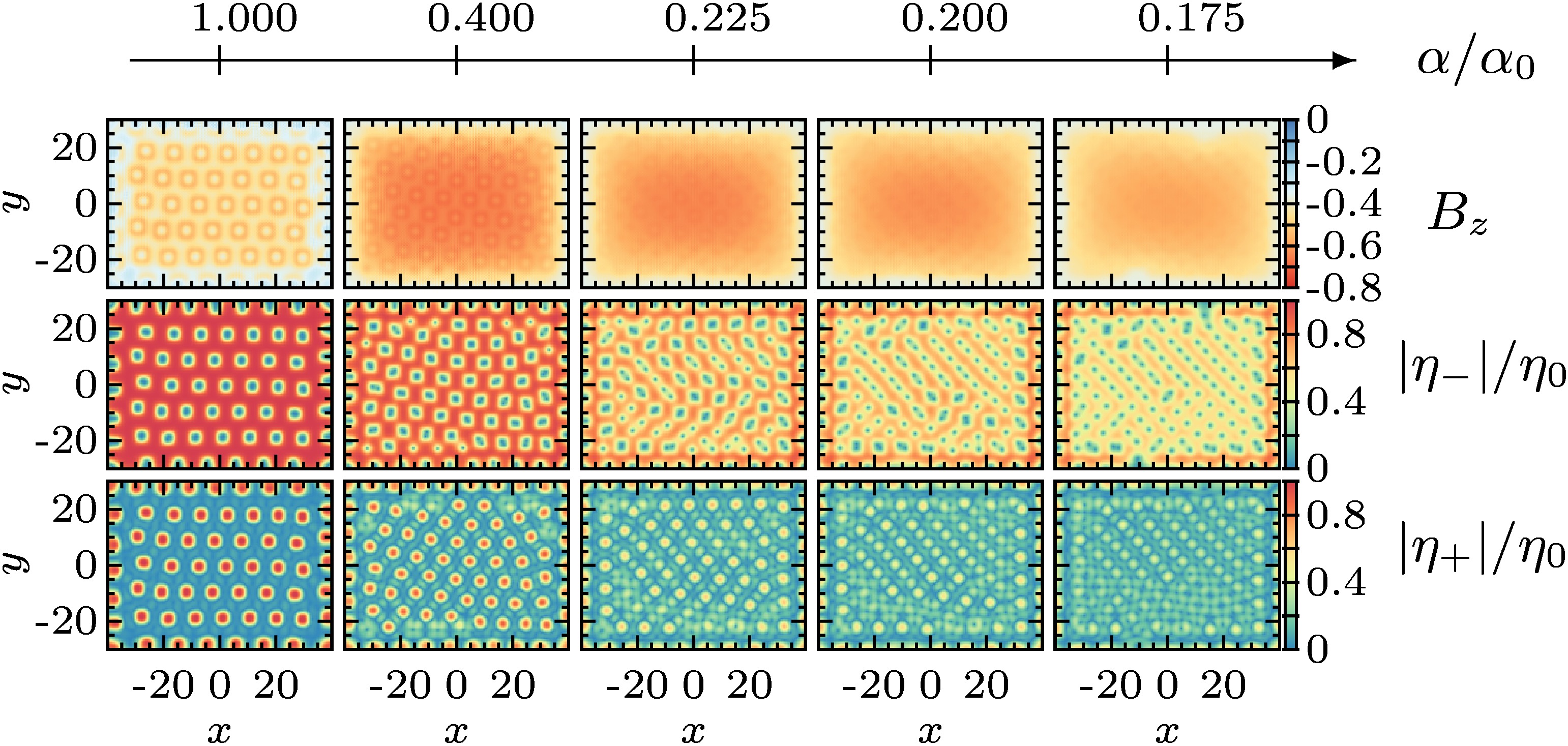}
\hss}
\caption{
(Color online) -- 
Simulation in an external field $\Bext<0$ for the parameters $g=0.3$, 
$\nu=0.3$, and $\chi=0.1$. The different lines display $B_z$, $|\eta_-|$, and 
$|\eta_+|$, respectively. Here, the external field is fixed and the prefactor 
of the quadratic term is varied, while other coefficients are kept fixed. 
This is equivalent to varying the temperature and getting closer to $\Hc{2}$. 
The first panel corresponds to the last panel of \Figref{Fig:RampUp1}. Decreasing 
$\alpha/\alpha_0$ decreases the total density. 
Close enough to $\Hc{2}$, the double-quanta vortices start to break apart. 
Eventually, very close to $\Hc{2}$ only single-quanta vortices subsist, 
and they arrange themselves in a square lattice.
}
\label{Fig:Split}
\end{figure*}

To investigate the properties near $\Hc{2}$, we need to slightly modify our 
parametrization of the theory, formulating it in a manner that is more convenient 
for numerical purposes
\footnote{
Approaching $\Hc{2}$ requires large fields, which can make numerical investigations 
difficult. Indeed, when applying higher fields, the domain is populated by more and 
more vortices. As a result, the complex fields have a larger and larger winding number 
at the boundary. Thus, in order to preserve reasonable accuracy (number of boundary 
points per winding number), one would need to refine the mesh, which would result in 
a dramatic slowdown of the numerics. Instead, horizontal displacement in the $(H,T)$-phase 
diagram allows us to approach $\Hc{2}$ without the special need to refine the mesh.
}.
The idea is that instead of approaching the upper critical field by varying $H$ at 
fixed $T$ in the $(H,T)$-phase diagram, the physics near $\Hc{2}$ can be found by 
varying $T$ at fixed $H$. In mean-field theory, \Eqref{Eq:FreeEnergy}, the 
temperature dependence is absorbed by setting the scales of the problem (here temperature 
refers to the temperature parameter of the non-fluctuating mean-field theory). We restore 
the parametrization of the temperature dependence by having the prefactor of the quadratic 
terms in \Eqref{Eq:FreeEnergy}: $\alpha(\tilde{T})=1-\tilde{T}$. Thus $\tilde{T}=1$ 
corresponds to the destruction of the superconducting state in zero field. Decreasing 
the value of the parameter $\alpha$ will thus decrease the superconducting density and 
push the system toward $\Hc{2}$.

Starting in the Meissner state with $\alpha=1$, the external field is gradually 
increased. The resulting magnetization process, similar to that displayed in 
\Figref{Fig:RampUp1}, produces a lattice of double-quanta vortices. Once the 
lattice is established, the applied field is fixed and the parameter $\alpha$ 
is sequentially decreased from $1$ to $0$ (in steps of $2.5\times10^{-2}$) and 
the energy is minimized at each step. 
Figure~\ref{Fig:Split} shows the evolution of a vortex lattice when decreasing 
$\alpha$ towards $\Hc{2}$. The system exhibits a lattice of double-quanta vortices 
for a rather wide range of temperatures. When getting closer to $\Hc{2}$, the lattice 
starts to deform and the double-quanta vortices split into single-quanta vortices. 
Eventually, the entire lattice of double-quanta vortices has dissociated into a 
structure of single-quanta vortices. Because finite-size effects become important 
together with a longer equilibration time, it becomes very difficult to form a 
fully ordered state. Thus, it is difficult to rigorously characterize such 
a lattice structure (see the discussion in Sec.~\ref{Sec:Numerics}). However, we can 
infer that our results, together with the earlier results based on lowest Landau-levels 
calculations \cite{Agterberg:98,Agterberg:98a}, point towards a transition to a 
lattice of a single-quanta vortices. Structures obtained by lowest Landau-levels 
calculations near $\Hc{2}$ are square lattices of the single-quanta vortices 
\cite{Agterberg:98,Agterberg:98a}.

The difference with the previously discussed scenario, is that our results indicate 
that square lattices of single-quanta vortices should transform into a hexagonal 
lattice of double-quanta vortices. The latter is robust and survives to large negative 
values of $\Bext<0$. Only in close vicinity to the upper critical field $\Hc{2}$ will 
double-quanta vortices dissociate. Note also that in the crossover region, single- 
and double-quanta vortices coexist, and there is a tendency to form vortex stripes.


\section{Chirality inversion and the role of domain-walls}
\label{Sec:Chirality-switching}

The discrete degeneracy of the chiral ground state is lifted by an external field 
and thus the magnetization processes should be different from that previously discussed, 
when the applied field is parallel to the ${\bf c}$-axis. 
Magnetization processes when $\Bext>0$ implies that the system can be in metastable states 
which are not energetically optimal. Indeed, the Meissner state with an initial chirality 
that does not correspond to the optimal direction of the applied field is not the one with 
the lowest energy.
Domain walls are natural topological excitations that interpolate between two ground 
states. In general, they are expected to form via a Kibble-Zurek-like mechanism 
\cite{Vadimov.Silaev:13}, but they could also play a role in the magnetization process 
where the starting state is not the optimal one in an external field. Various aspects 
of domain-wall properties during magnetization processes have been studied in 
\cite{Matsunaga.Ichioka.ea:04,Ichioka.Matsunaga.ea:05}. After briefly reviewing their 
elementary properties, we investigate the magnetization processes when $\Bext>0$. This 
magnetization process is actually much richer than that taking place when the starting
state is an optimal one in a given external field. Indeed, for non-optimal starting 
states, we will discover that the magnetization process involves chirality inversion 
processes, the details of which will be sensitive to the parameters of the theory.

\subsection{Domain walls}
\label{Sec:DW}

A domain wall is a field configuration that interpolates, for example, between 
$(|\eta_+|,|\eta_-|)=(1,0)$ and $(|\eta_+|,|\eta_-|)=(0,1)$. Note that there are 
two inequivalent ways of having such a configuration, with differing corresponding 
domain walls. The two inequivalent ways may be illustrated by
\SubAlign{Eq:DW}{
&\mathrm{DW_{I\phantom{I}}}:~~~~
	(-1,0) \longleftarrow (\eta_+,\eta_-) \longrightarrow (0,1) \\
&\mathrm{DW_{II}}:~~~~
	(\phantom{-}1,0) \longleftarrow (\eta_+,\eta_-) \longrightarrow (0,1)\,.
}
It is easily realized that the two domain wall configurations cannot be transformed 
into each other by gauge transformations, from which they are physically distinguishable.
Note that the energy cost of a domain wall also depends on its relative orientation 
with respect to the crystal axis. Depending on the orientation of the domain 
wall, one of the two possible domain walls is favored. This was discussed in 
detail in Ref.~\cite{Bouhon.Sigrist:10}.

\begin{figure}[!b]
\hbox to \linewidth{ \hss
\includegraphics[width=0.9\linewidth]{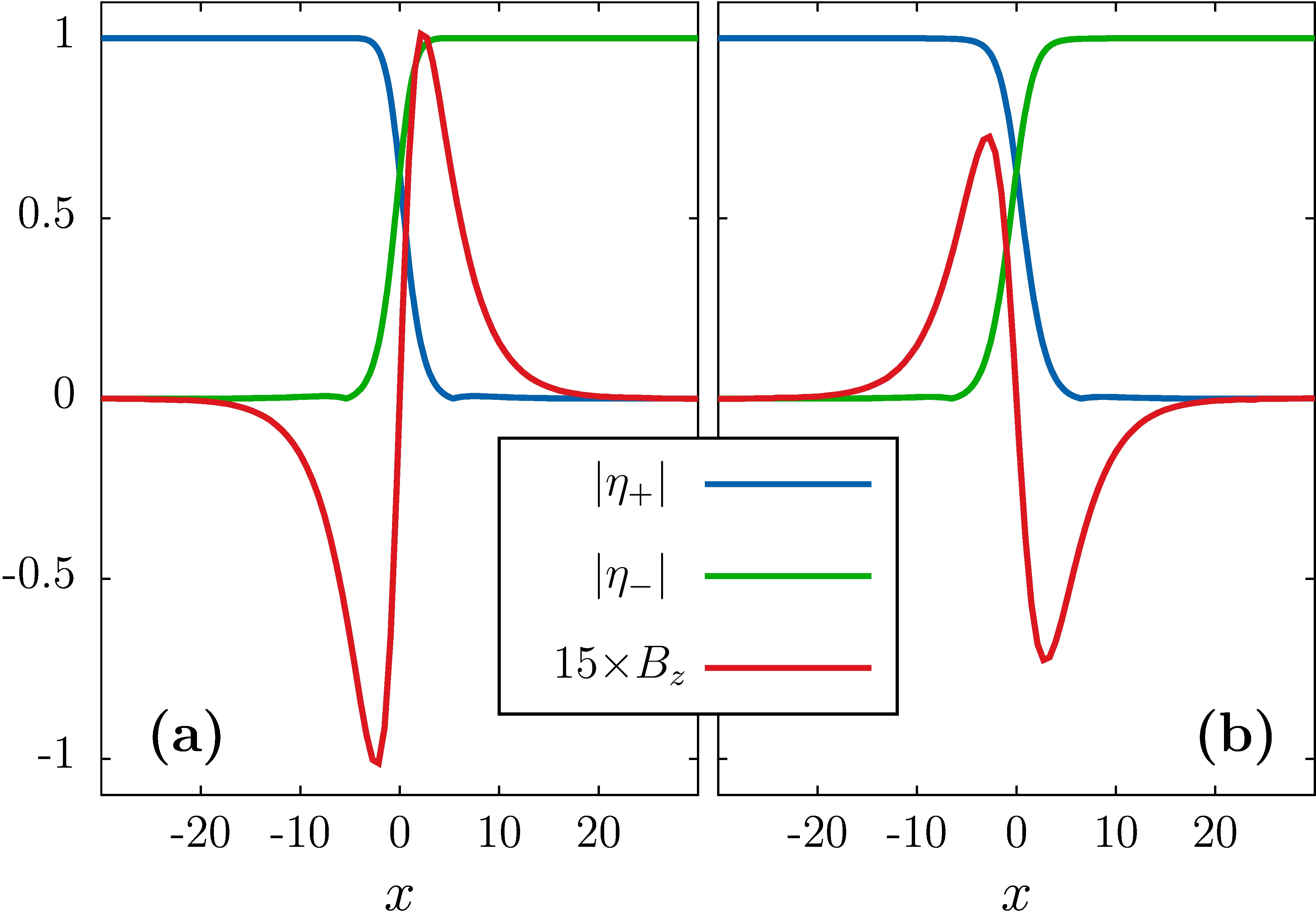}
\hss}
\caption{
(Color online) -- 
The two possible kinds of domain walls interpolating between 
$(|\eta_+|,|\eta_-|)=(1,0)$ and $(|\eta_+|,|\eta_-|)=(0,1)$, for the parameters 
$g=0.3$ and $\nu=-0.5$. Panels (a) and (b) show $\mathrm{DW_{I}}$ 
and $\mathrm{DW_{II}}$ domain walls, respectively. 
Their density profiles are very similar, but they differ from their real and 
imaginary parts and they cannot be transformed into each other. Although the density 
profiles are quite close to each other, the two domain walls have different energies 
and also different magnetic field due the difference in relative phases.
}
\label{Fig:DomainWalls}
\end{figure}

Figure~\ref{Fig:DomainWalls} displays the typical domain wall solution in 
chiral $p$-wave superconductors. The magnetic signatures of the two types of 
domain walls \Eqref{Eq:DW} are different, and they have different energies. 
Due to partial currents in different chiralities, the domain walls have 
longitudinal currents associated with them, and hence they carry a magnetic 
field, as can be seen from \Figref{Fig:DomainWalls}. 
Conversely, since the domain walls support longitudinal currents, an 
external applied field will produce a Lorentz force that should induce 
motion of the domain wall. 
In other words, when the degeneracy between ground states is lifted by an external 
field, the domain wall should move to increase the region of optimal ground state. 
Thus, we expect domain walls to be involved in the magnetization processes when 
the external field is not optimally oriented.

\vspace{0.5cm}
\subsection{Chirality inversion in an external field}

\begin{figure*}[!htb]
\hbox to \linewidth{ \hss
\includegraphics[width=.9\linewidth]{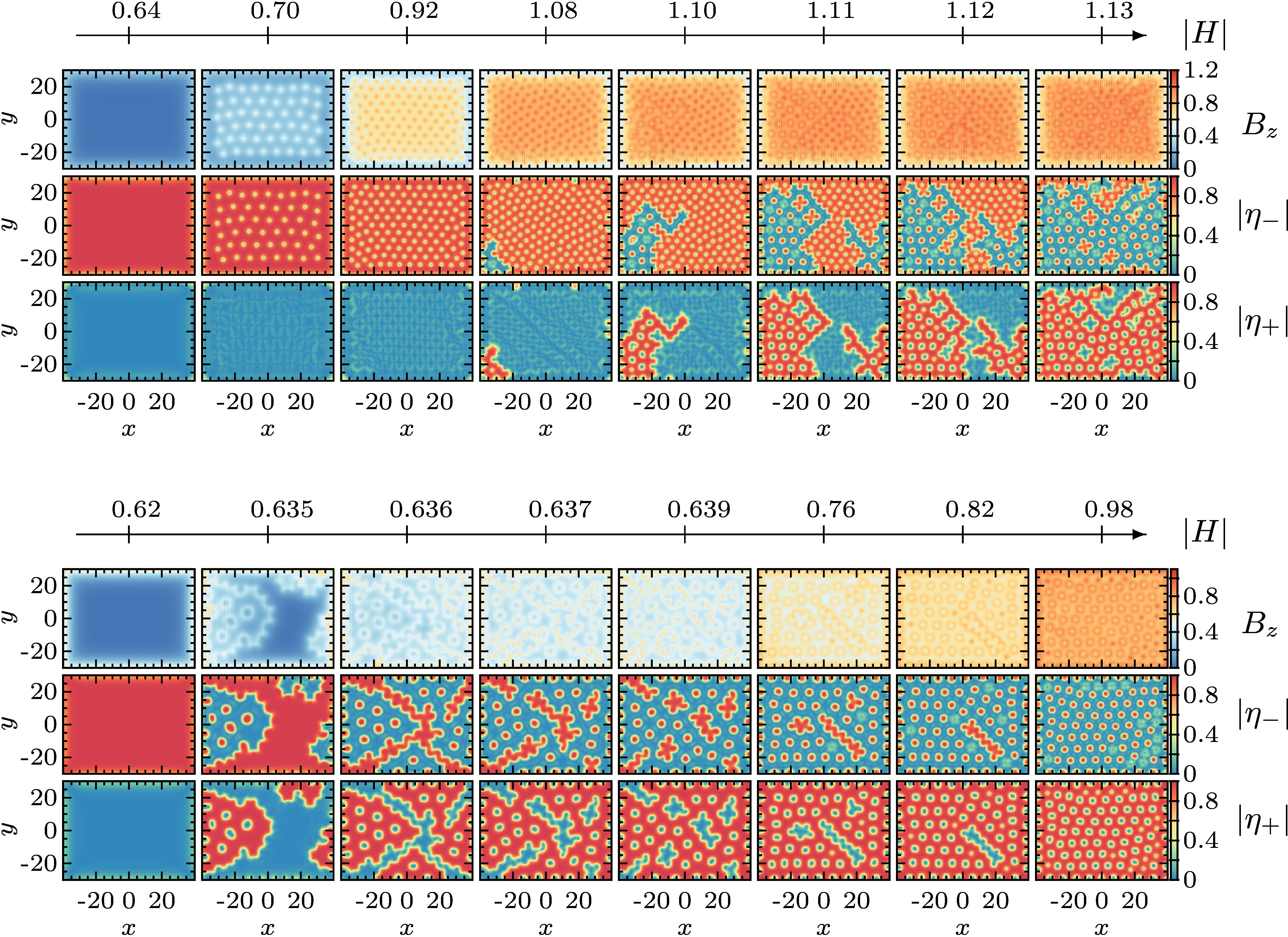}
\hss}
\caption{
(Color online) -- 
Simulations in an external field $\Bext>0$ for the parameters $g=0.3$ and $\nu=0.15$. 
The different lines display $B_z$, $|\eta_-|$, and $|\eta_+|$, respectively. Here the 
orientation of the external field is such that it produces topological defects with 
higher energy: single-quanta (singular) vortices are favored over skyrmions.
Given the direction of the applied field the initial chirality is non-optimal.
The parameter of the edge currents are $\chi=1$ for the top panel and $\chi=10$ 
for the lower one.
In the top panel, upon increasing the external field, single-quantum vortices enter 
and organize as a lattice. At elevated fields, a domain wall starts entering and 
``reverting'' the chirality. 
In the lower panel, the domain wall starts entering the domain and switching the 
chirality before any vortex entry. The domain walls here also carry vorticity, as it 
becomes energetically beneficial to place vortices there.
In both cases, behind the domain wall, the optimal chirality double-quanta skyrmions 
are the lowest-energy excitation. Eventually, mostly double-quanta skyrmions occupy the 
domain and should turn into a lattice of skyrmions.
}
\label{Fig:RampUp2}
\end{figure*}

Domain walls are the topological excitations that are involved in processes that revert 
the chirality. For an applied field parallel to the $c$-axis ($\Bext>0$), the ground 
state $(\eta_+,\eta_-)=(0,1)$ is not the optimal one. Thus, two isolated single-quanta
vortices have lower energy than a double-quantum vortex. That is, $(n_+,n_-)=(+3,+1)$ 
vortices have a smaller lower critical field than the double-quanta  $(n_+,n_-)=(+4,+2)$ 
vortices, i. e.  $\Hc{1}^{(n_-=+1)}<\Hc{1}^{(n_-=+2)}$. The top panel in \Figref{Fig:RampUp2} 
illustrates this, and only single-quanta vortices enter and organize as a 
lattice. Note that the field for the first vortex entry in this case is higher than 
for anti-parallel field, since the $\Hc{1}^{(n_-=+1)}>\Hc{1}^{(n_-=-1)}$. Therefore, 
single-quanta vortices enter and arrange themselves as a lattice in low field. An 
interesting process occurs in higher fields. Since the ground state 
$(\eta_+,\eta_-)=(0,1)$ is not optimal for that direction of the external field, 
the optimal case would thus actually be to have the opposite chirality. For rather 
large fields, we see that the systems starts to ``reverse'' its chirality. By nucleating 
a domain wall that propagates from the boundaries, the system is able to switch to optimal 
chirality, given the orientation of the external field. 
While the domain wall propagates in the bulk, it ``absorbs'' the single-quanta vortices 
and ``converts'' them into double-quanta vortices in the optimal chirality. Eventually, 
mostly double-quanta skyrmions occupy the domain and should turn into a lattice of 
skyrmions.

During the process of chirality inversion, various kinds of vortices carrying 
different numbers of flux quanta can coexist. For instance, there are a few single-quanta
vortices that are trapped between double-quanta vortices. They cannot always pair with 
other single-quanta vortices, as this would imply moving through the background of other 
double-quanta vortices. Such trajectories can be energetically unfavored. Similarly, 
skyrmions carrying more than two flux quanta are also formed and persist since these 
are metastable solutions. Their decay into double-quanta vortices can be triggered by the 
pressure that is exerted by the surrounding double-quanta skyrmions. We find that the 
skyrmions carrying high magnetic flux are eventually destroyed by an increasing field.

Another possible scenario for the magnetization process with external field parallel 
to the $c$-axis, is displayed in the lower panel of \Figref{Fig:RampUp2}. Typically 
for small geometries, or for a strong barrier to vortex entries, it may be beneficial 
to produce a domain wall at field below the lower critical field, and thus switch to 
the optimal chirality prior to any vortex entry. Besides the domain wall, the optimal 
chirality double-quanta skyrmions are the lowest energy excitations. The created domain 
walls are not the bare domain walls discussed above in Sec.~\ref{Sec:DW}, 
but rather domain walls `decorated' by vortices, such that they carry vorticity. 
As a result of the vorticity which is trapped on the domain walls, they cannot easily 
annihilate (with an anti-DW), so they create skyrmions with a large number of flux 
quanta (see detail of the mechanism of stabilization of domain walls by vortex decoration 
in \cite{Garaud.Babaev:12,Garaud.Babaev:15a}). At elevated fields, however, these decorated 
domain walls eventually decay when the system is compressed enough, leaving a lattice 
of double-quanta vortices (in addition to a few isolated single quanta); the optimal 
chirality has been restored. From now on, the behavior of the double-quanta vortex 
lattice is the same as that discussed in in Sec.~\ref{Sec:double-quanta-lattices}.
That is, further increasing the external field will drive the structural transition 
into a single-quanta square lattice close to $\Hc{2}$, accompanied by a density halving of 
the lattice.

We have found that a magnetic field anti-aligned with chirality should trigger a chirality 
inversion process by propagation of domain walls ``decorated'' with vortices inside the 
domain. We report two possibilities for such an inversion process, namely that domain wall 
penetration occurs either before or after penetration of single quanta vortices. 
Weak edge currents promote early entry of single-quanta vortices prior to the domain 
wall penetration and chirality inversion process. Strong edge currents, on the other 
hand, delay entry of single-quanta vortices compared to the domain wall. In that case, 
the restoration of the optimal chiral state is much faster.
Note that which of the two scenarios is realized depends not only on the strength of the 
edge currents, but also on the size and shape of the domain that is considered.

\section*{Conclusion}

In this paper, we have considered the problem of magnetization of a finite 
superconducting sample in the framework of a standard Ginzburg-Landau model 
for chiral $p$-wave superconductors that is often invoked to describe Sr$_2$RuO$_4$. 
At magnetic fields close to $\Hc{2}$, there is a tendency towards formation of 
a square lattice of single-quantum vortices, in agreement with earlier calculations 
\cite{Agterberg:98,Agterberg:98a} and experimental observations 
\cite{Aegerter.Lloyd.ea:98,Riseman.Kealey.ea:98,Ray.Gibbs.ea:14}. However, we 
find that, at least at mean-field level in the Ginzburg-Landau  model, the square 
lattice exists only very close to $\Hc{2}$ and transforms into a hexagonal lattice 
of double-quanta vortices slightly below $\Hc{2}$. This double-quanta hexagonal 
vortex lattice dominates the phase diagram of the model in question. 
In contrast to the Eilenberger theory-based calculations in 
Ref.~\cite{Ichioka.Machida.ea:12}, in our calculations the double-quanta vortex 
lattice persists down to the lowest fields. Double-quanta vortex formation has 
also been reported in simulations of mesoscopic samples in external fields 
\cite{FernandezBecerra.Sardella.ea:16}.

Different chiralities are known to have different lower critical fields $\Hc{1}$.  
For the chirality with larger $\Hc{1}$, we have found metastable hexagonal vortex 
lattices of single-quanta vortices in low magnetic fields. The metastable single-quanta 
vortex lattices transform into a stable double-quanta vortex lattice at elevated fields 
via a set of complicated metastable states that involve the creation and growth of domain 
walls decorated by vortices. These metastable configurations have characteristic magnetic 
field signatures that should be detectable by scanning SQUID and Hall probes or decoration.

Although our results are inconsistent with the current experimental data on Sr$_2$RuO$_4$ 
\cite{Bjornsson.Maeno.ea:05,Hicks.Kirtley.ea:10,Ray.Gibbs.ea:14}, they do not rule 
out $p$-wave superconductivity in this material. Rather, our results present evidence against a 
class of minimal models. This magnetization picture can be used as a ``smoking gun'' hallmark 
of chiral $p$-wave superconductivity that is searched for in other materials.

\begin{acknowledgments}
We acknowledge fruitful discussions with Mihail Silaev.
The work of JG and EB was supported by the Swedish Research Council Grant  
No. 642-2013-7837 and the Goran Gustafsson Foundation. 
The work of AS was supported by the Research Council of Norway Grants 
No. 205591/V20 and No. 216700/F20, as well as European Science Foundation 
COST Action MPI1201. 
The computations were performed on resources provided by the 
Swedish National Infrastructure for Computing (SNIC) at the National 
Supercomputer Center in Link\"oping, Sweden.
\end{acknowledgments}

\vspace{0.25cm}
\appendix
\section{Vortices in terms of \texorpdfstring{$\eta_{x,y}$}{etaxy} components}
\label{App:VorticesPXPY}

\begin{figure}[!htb]
\hbox to \linewidth{ \hss
\includegraphics[width=0.95\linewidth]{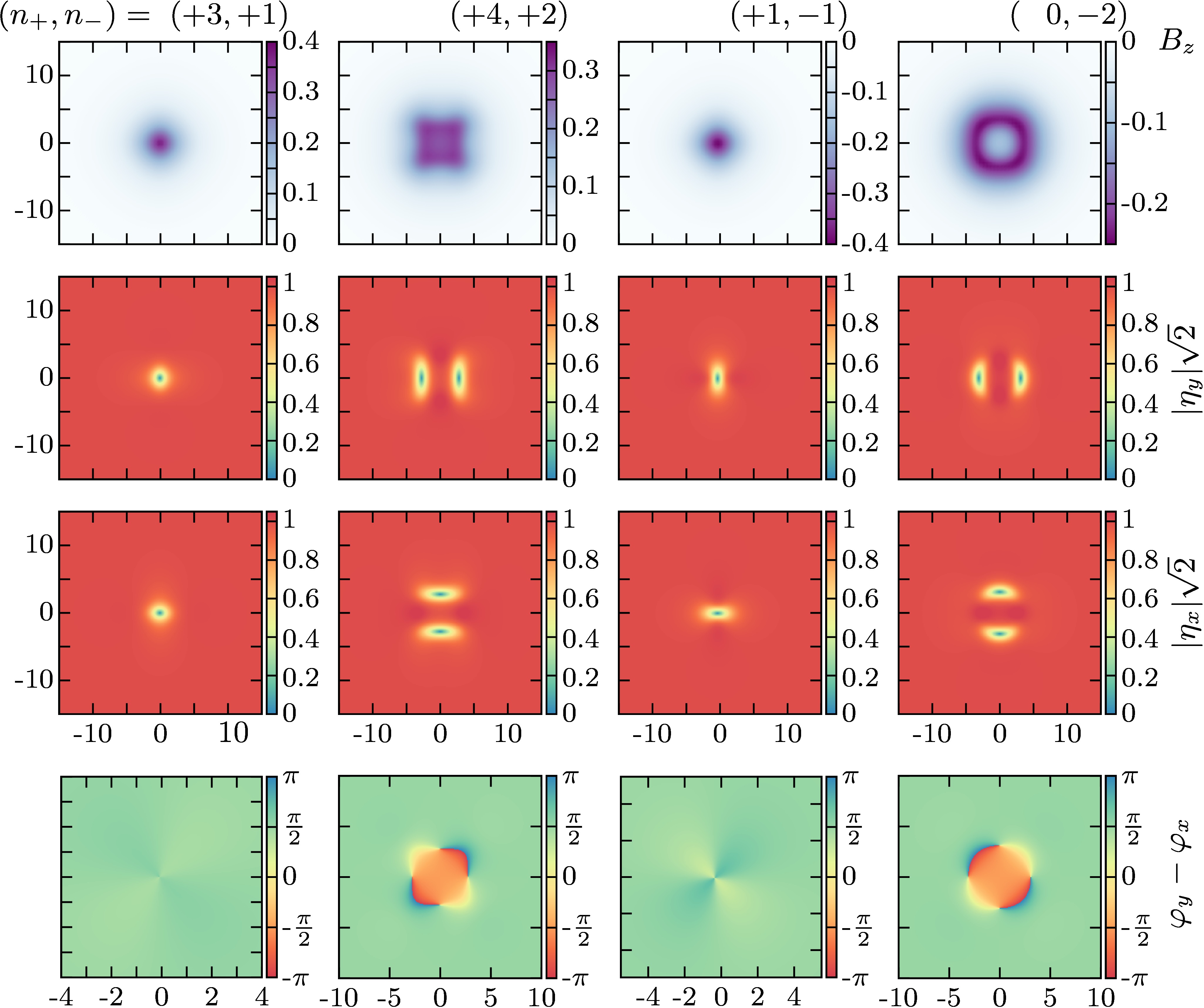}
\hss}
\caption{
(Color online) -- 
Vortex states for the parameters $g=0.3$ and $\nu=0.2$. The first row shows the magnetic 
field $\B$, while the second and third row display $\eta_y$ and $\eta_x$, respectively. 
The fourth line shows the relative phase $\varphi_y-\varphi_x$ between $\eta_x$ and 
$\eta_y$. The first two columns show single- and double-quanta vortices, respectively, 
with $B_z>0$, while the third and fourth columns display single- and double-quanta vortices 
with $B_z<0$. 
}
\label{Fig:VorticesPXPY}
\end{figure}

It is instructive to consider the configurations displayed in \Figref{Fig:Vortices}
in the $(\eta_x,\eta_y)$ order parameter basis, as is done in \Figref{Fig:VorticesPXPY}. 
There, the two inequivalent ground states have equal density and are distinguished by 
the relative phase $\varphi_y-\varphi_x$ (between $\eta_x$ and $\eta_y$) being $\pm\pi/2$. 
Again it is quite clear that opposite vorticities give different structures of the cores.
The parametrization in terms of $(\eta_x,\eta_y)$ sheds new light on how to interpret 
the double-quanta vortices. Since in this parametrization both $\eta_x$ and $\eta_y$ have 
non-zero ground-state density, both components can have non-zero (asymptotic) winding and 
thus contribute equally to screening of the magnetic field. A vortex within each component 
can be attributed half of a flux quantum, and a bound state of a half-quantum vortex in 
each component constitutes a single quantum vortex. 
%

%
\end{document}